\documentstyle[epsf,12pt]{article}
\newcommand{\beq}{\begin{equation}}
\newcommand{\eeq}{\end{equation}}
\newcommand{\beqa}{\begin{eqnarray}}
\newcommand{\eeqa}{\end{eqnarray}}

\newcommand{\bma}{\left( \begin {array}}
\newcommand{\ema}{\end {array} \right)}
\newcommand{\bfig}{\begin{figure}}
\newcommand{\efig}{\end{figure}}
\newcommand{\bc}{\begin{center}}
\newcommand{\ec}{\end{center}}
\newcommand{\pslash}{\kern 0.2 em p\kern -0.45em /}
\newcommand{\sla}[1]{\kern 0.2 em #1\kern -0.45em /}

\begin{document}
\setcounter{page}{0}
\thispagestyle{empty}
\hspace*{13.0cm}                                  CERN-TH.6900/93\\
\hspace*{13.0cm}                                  WU-B 93-19\\
\bc
{\Large\bf THE PION FORM FACTOR: \\
           SUDAKOV SUPPRESSIONS AND\\
           INTRINSIC TRANSVERSE MOMENTUM \\}
\vspace{1cm}
\ec
\bc
R. Jakob
\footnote{Supported by the Deutsche Forschungsgemeinschaft}
\footnote{E-mail: rainer@wpts0.physik.uni-wuppertal.de}
\\
Fachbereich Physik, Universit\"{a}t Wuppertal, Gau\ss strasse 20,\\
Postfach 10 01 27, D-5600 Wuppertal 1, Germany\\[0.3 cm]
and\\[0.3 cm]
P. Kroll
\footnote{On leave of absence from Fachbereich Physik,
Universit\"at Wuppertal, Germany
\newline
\newline
\newline
\normalsize{
CERN-TH.6900/93
\newline
WU-B 93-19
\newline
May 1993}}\\
Theory Division, CERN, CH-1211 Geneva 23, Switzerland
\ec
\bc
                    {\bf Abstract}
\ec
It is demonstrated that any attempt to calculate the perturbative QCD
contribution to the pion form factor requires the inclusion of
intrinsic transverse momentum besides Sudakov form factors. For
momentum transfers of the order of a few GeV the intrinsic transverse
momentum leads to a substantial suppression of the perturbative QCD
contribution.
\newpage
There is general agreement that perturbative QCD in the framework of the
hard-scattering picture \cite{Lep:80} is the correct description of exclusive
reactions at asymptotically large momentum transfer. However, the
applicability of this approach at experimentally accessible momentum
transfers, typically a few GeV, was questioned \cite{Isg:89,Rad:91}. It
was asserted that in the few GeV region the hard-scattering picture
accumulates strong contributions from the soft end-point regions, rendering
the perturbation calculation inconsistent. Recently, this statement was
disproved by Sterman and
collaborators \cite{Bot:89,LiS:92,Li:92}. The essential point
of their work is that the customarily neglected transverse momenta
of the quarks as well as the Sudakov corrections are
taken into account. Because of that the perturbative QCD
contribution becomes self-consistent, even for momentum transfers
as low as a few GeV, and seem to dominate form factors and
cross-sections. For the electromagnetic form factor of the pion, the
simplest exclusive quantity, it has been shown \cite{LiS:92} that
perturbative QCD can readily be applied for momentum transfers larger
than $20\Lambda_{QCD}$. It
should be noted that the analyses of
\cite{Bot:89,LiS:92,Li:92} involve no phenomenological parameter.\\
However, it seems to us that an important element is missed in the
analyses of \cite{LiS:92,Li:92}, namely the intrinsic transverse
momentum dependence of the hadronic wave function. As we are going to
demonstrate for the case of the pion form factor, the inclusion of that
$k_\perp$-dependence substantially suppresses the contribution from
perturbative QCD. Although self-consistently calculated, this
contribution then seems to be too small with respect to the data. We
therefore suspect that there are sizable soft contributions to
the pion form factor in the few GeV region, a
conjecture already expressed
in \cite{Isg:89,Rad:91}.\\
In order to make our point of view clear it is useful to repeat the
essential steps in the derivation of the hard-scattering formula
for the pion electromagnetic form factor \cite{Lep:80}. The starting
point is the Drell-Yan formula \cite{Dre:70} in which the
pion form factor is expressed as an overlap of the initial-
and final-state light-cone wave functions
\beq
\label{eq:drell-yan}
F_\pi(Q^2) = \int \frac{dx_1 \, d^{\;\!2} k_\perp}{16 \pi^3}
\,\Psi^\ast (x_1,\vec k_\perp+x_2\vec q) \, \Psi(x_1,\vec k_\perp),
\eeq
where $Q^2=\vec q^{\;\!2}$ is the momentum
transfer, $x_1$ is the usual
longitudinal momentum fraction carried by the quark and $\vec k_\perp$ its
transverse momentum with respect to its parent hadron. The
momentum of the antiquark is characterized by $x_2=1-x_1$ and
$-\vec k_\perp$; $\Psi$ is the wave function of the valence Fock
state
\footnote{Strictly speaking $\Psi$ is the $S=0, L_z =0$ component of the
valence Fock state wave function.}. It
satisfies the normalization condition
\beq
\label{eq:probability}
\int \frac{dx_1 \, d^{\;\!2} k_\perp}{16 \pi^3}
\left| \Psi (x_1,\vec k_\perp) \right|^2 = P_{q \bar q}
\leq 1.
\eeq
Contributions from higher Fock states are neglected
in (\ref{eq:drell-yan}) since, at large momentum transfer, they
are suppressed by inverse powers of $Q^2$.
As the inspection of (\ref{eq:drell-yan}) reveals, the
large $Q$ behaviour of the pion form factor is controlled by the
tail of the wave function at large $\vec k_\perp$. The crucial
point is that the tail of the wave function  can be
calculated, within perturbative
QCD, from the soft part of the wave function by means of a homogeneous
Bethe-Salpeter equation. Absorbing the perturbative kernel of the
Bethe-Salpeter equation in a hard-scattering
amplitude $T_H$, expression (\ref{eq:drell-yan}) can be converted
into
\beq
\label{eq:pre-hs-Fpi}
F_\pi(Q^2)=
\int \frac{dx_1 \, d^{\;\!2} k_\perp}{16 \pi^3}
\int \frac{dy_1 \, d^{\;\!2} l_\perp}{16 \pi^3} \,
\,\Psi_0^\ast (y_1,\vec l_\perp)\, T_H (x_1,y_1,Q,\vec k_\perp,\vec l_\perp)
\, \Psi_0 (x_1,\vec k_\perp).
\eeq
Now $\Psi_0$ represents the soft part of the pion wave function, i.e. the
full wave function with the perturbative tail removed from it. The
perturbative tail of $\Psi$ behaves as $1/k_\perp^2$ for
large $k_\perp$ whereas $\Psi_0$ vanishes as $\sim 1/k_\perp^4$ or
faster. To lowest order in perturbation theory, the hard-scattering
amplitude is to be
calculated from the one-gluon-exchange diagrams. Keeping the transverse
momenta of the quarks, $\vec k_\perp$ and $\vec l_\perp$, only
in the gluon propagators, where they matter most
\footnote{The effect of the $k_\perp$-dependence in the fermion propagators
has been investigated by Li \cite{Li:92}; it leads
to a reduction of the results for $F_\pi$ by about $10\%$ for
$Q > 3\,{\rm GeV}$.}, the
hard-scattering amplitude reads
\beq
\label{eq:hs-amplitude}
T_H (x_1,y_1,Q,\vec k_\perp,\vec l_\perp) =
\frac{16 \pi \, \alpha_s(\mu) \, C_F}
{x_1 y_1 Q^2+(\vec k_\perp + \vec l_\perp)^2},
\eeq
where we have made use of the symmetry of the pion wave function under
the replacement $x_1 \leftrightarrow x_2$ ($C$ invariance). In
eq. (\ref{eq:hs-amplitude}) $\alpha_s$ is the usual strong coupling
constant to be evaluated at a renormalization scale $\mu$ and
$C_F \,(=4/3)$ is the colour factor.\\
At large $Q$ one may neglect the $k_\perp$- and $l_\perp$-dependence
in the gluon propagator as well; $T_H$ can then be pulled out of
the transverse momentum integrals, and these
integrations apply only to the
wave functions. Defining distribution amplitudes (DA) by
\beq
\label{eq:DAdef}
\frac{f_\pi}{2 \sqrt{6}} \,\, \phi (x_1,\mu) =
\int \frac{d^{\;\!2} k_\perp}{16 \pi^3}\,  \Psi_0 (x_1,\vec k_\perp),
\eeq
one arrives at the celebrated hard-scattering formula for the pion
form factor
\beq
\label{eq:hs-Fpi}
{F_\pi}^{HSP}(Q^2)=\frac{{f_\pi}^2}{24} \int dx_1 \, dy_1
\, \phi^\ast(y_1,\mu) \, T_H(x_1,y_1,Q,\mu) \, \phi(x_1,\mu),
\eeq
which is valid for $Q \to \infty$.\\
The DA is defined such that
\beq
\label{eq:DAnorm}
\int_0^1 dx_1 \, \phi(x_1,\mu) = 1.
\eeq
An immediate consequence of the definitions (\ref{eq:DAdef}) and
(\ref{eq:DAnorm}) is that the constraint from the $\pi \to \mu\nu $ decay
\cite{Lep:83} is automatically satisfied:
\beq
\label{eq:pi-munu}
\frac{f_\pi}{2 \sqrt{6}} = \int \frac{dx_1 \, d^{\;\!2} k_\perp}{16 \pi^3}
\,\Psi_0 (x_1,\vec k_\perp),
\eeq
where $f_\pi \,(=133\, {\rm MeV})$ is the usual $\pi$ decay constant. The
integral in (\ref{eq:DAdef}) has to be cut off at a scale of
order $Q$, which leads to a
very mild dependence of the DA on the renormalization
scale (QCD evolution). For most applications of
perturbative QCD to exclusive reactions, the $\mu$-dependence of the
DAs can be ignored since one is only interested in a rather limited
region of $Q$.\\
The neglect of the transverse momentum dependence
in (\ref{eq:hs-amplitude}) is a bad approximation in the end-point
regions where one of the momentum fractions, $x_i$ or $y_i$, $i=1,2$, tends
to zero. This approximation is after all responsible
for the theoretical
inconsistencies mentioned in the introduction. Li and
Sterman \cite{LiS:92} have suggested to retain the transverse momentum
dependence of $T_H$. Moreover, Sudakov corrections, suppressing the
contributions from the dangerous soft regions, ought to be taken
into account. In order to include these corrections it is
advantageous to reexpress eq. (\ref{eq:pre-hs-Fpi}) in
terms of the Fourier transform variable $\vec b$ in the
transverse configuration space
\footnote{Since $T_H$ depends only on $\vec k_\perp + \vec l_\perp$ there
is merely one independent variable in the transverse configuration space.}
\beqa
\label{eq:ft-Fpi}
{F_\pi}^{pert}(Q^2)&=&
\int_0^1\! \frac{dx_1 \, dy_1}{(4 \pi)^2}
\int_{-\infty}^\infty \!d^{\;\!2}b
\,\hat{\Psi}_0^\ast (y_1,\vec b,w)\,\hat{T}_H (x_1,y_1,Q,b,t)
\,\hat{\Psi}_0 (x_1,-\vec b,w) \nonumber \\
& &\qquad\qquad\qquad\qquad\qquad
\times\,\exp\left[-S(x_1,y_1,Q,b,t)\right]
\eeqa
where the Fourier transform of a function $f=f(\vec k_\perp)$ is
defined by
\beq
\label{eq:Fourier-transformation}
\hat f(\vec b)=\frac{1}{(2\pi)^2}
\int \!d^{\;\!2}k_\perp
\,e^{-i\vec b\cdot \vec k_\perp}
\,f (\vec k_\perp).
\eeq
The Fourier-transformed hard-scattering amplitude reads
\beq
\label{eq:ft-hs-amplitude}
\hat{T}_H (x_1,y_1,Q,b,t) =
8\, \alpha_s(t) \, C_F \, K_o(\sqrt{x_1 y_1}\,Q\,b),
\eeq
where $K_0$ is the modified Bessel function of order zero and $t$ is
the largest mass scale appearing in $\hat{T}_H$:
\beq
\label{eq:als-arg}
t= {\rm Max}(\sqrt{x_1 y_1}\,Q,w=1/b).
\eeq
The factor $\exp \left[-S\right]$ in
(\ref{eq:ft-Fpi}), termed the Sudakov factor, comprises the radiative
corrections. The
function $S$ is given by \cite{Bot:89}
\beq
\label{eq:sudakov}
S(x_1,y_1,Q,b,t)=\sum_{i=1}^2
\left[s(x_i,Q,b)+s(y_i,Q,b)\right]
-\frac{2}{\beta_1} \ln \frac{\ln (t/\Lambda_{QCD})}
{\ln (1/b \Lambda_{QCD})}.
\eeq
The last term in (\ref{eq:sudakov}) arises from the application of the
renormalization group. The lengthy expression for the Sudakov
exponent $s$, which includes all leading and next-to-leading
logarithms, is given explicitly in \cite{LiS:92}. The most important
term in it is the double logarithm
\beq
\label{eq:double-log}
\frac{2}{3\beta_1} \ln \frac{\xi Q}{\sqrt{2} \Lambda_{QCD}}
\ln \frac{\ln (\xi Q/\sqrt{2} \Lambda_{QCD})}{\ln (1/b \Lambda_{QCD})},
\eeq
where $\xi$ is one of the fractions, $x_i$ or $y_i$, and
$\beta_1= (33-2n_f)/12$. For small $b$, i.e. at small transverse
separation of quark and antiquark, there is no suppression from the
Sudakov factor\footnote{Actually
the Sudakov factor includes a small enhancement
in the low-$b$ region. Li and Sterman \cite{LiS:92} neglect this
enhancement and set any factor $\exp\left[-s(\xi,b,Q)\right] $ to
unity in this region and whenever $\xi < \sqrt{2}/bQ $. We
apply this recipe too.}. As
$b$ increases the Sudakov factor decreases, reaching zero at
$b=1/\Lambda_{QCD}$. For $b$ larger than $1/\Lambda_{QCD}$ the Sudakov
factor is set to zero. Owing to this cut-off the singularity
of $\alpha_s$ is avoided without introducing a phenomenological
parameter (e.g. a gluon mass). For $Q \to \infty$ the Sudakov
factor dumps any contribution except those from configurations with
small quark-antiquark separation. In other words, the hard-scattering
contributions dominate the pion form factor asymptotically.\\
Li and Sterman have explored the improved hard-scattering
formula (\ref{eq:ft-Fpi}) on the basis of customary
wave functions, neglecting the QCD evolution and the intrinsic transverse
momentum dependence of the wave functions. Their numerical studies
have revealed
that the modified perturbative approach is self-consistent for
$Q>20\Lambda_{QCD}$ in the sense that less than, say, $50\%$ of the
result is generated by soft gluon exchange $(\alpha_s > 0.7)$. In
the few GeV region the values for $F_\pi$ as obtained by Li and Sterman
are somewhat smaller than those provided by the hard-scattering formula
(\ref{eq:hs-Fpi}) and are, perhaps, smaller than the experimental
values \cite{Beb:76}. The data may however suffer from large systematic
errors \cite{Car:90}. This prevents us from giving a definite
conclusion about the agreement or disagreement between data and theory.\\
The approach proposed by Sterman and
collaborators \cite{Bot:89,LiS:92,Li:92}
certainly constitutes an enormous progress in our understanding of
exclusive reactions at large momentum transfer. We believe, however, that
in any practical application of that approach one has to allow for an
intrinsic transverse momentum dependence of the hadronic wave
function, although, admittedly, this requires a new phenomenological element
in the calculation. In order to demonstrate the importance of the
intrinsic transverse momentum for the case of the pion form factor, we
need realistic pion wave functions.\\ \\
\underline{\it The pion wave function:} In accordance with
(\ref{eq:DAdef}), (\ref{eq:DAnorm}) and (\ref{eq:pi-munu}) we
write the soft part of the valence quark Fock state wave function as
\beq
\label{eq:wvfct}
\Psi_0 (x_1,\vec k_\perp) = \frac{f_\pi}{2 \sqrt{6}}
\,\phi(x_1) \,\Sigma(x_1,\vec k_\perp),
\eeq
the function $\Sigma$ being normalized in such a way that
\beq
\label{eq:Sigmanorm}
\int \frac{d^{\;\!2} k_\perp}{16 \pi^3} \,\Sigma (x_1,\vec k_\perp) =1.
\eeq
The probability of the valence quark Fock state is given by
(\ref{eq:probability})
\footnote{The use of the soft part $\Psi_0$ of the wave function in
(\ref{eq:probability}) instead of the full one, entails an
insignificant $\cal O (\alpha_s)$ error.}. A
further constraint of the wave function comes from the charge
radius of the pion. The (transverse) radius of the valence Fock state
should be smaller than (or at best equal to) the full
radius. Or, in terms of transverse
momentum, we have a lower limit on the mean square transverse momentum
that is defined by
\beq
\label{eq:expect-k}
\langle{k_\perp}^2\rangle=
\frac{{f_\pi}^2}{24 \, P_{q \bar q}}
\int \frac{dx_1 \, d^{\;\!2}k_\perp}{16 \pi^3}
\,k_\perp^2 \,|\phi(x_1)|^2 \,|\Sigma(x_1,\vec k_\perp)|^2.
\eeq
Actually, the root mean square transverse
momentum (r.m.s.), $\langle k_\perp^2 \rangle^{1/2}$, should
be larger than $300 \,{\rm MeV}$
\footnote{Lepage et al. \cite{Lep:83} have
proposed to determine the radius, and
in turn the r.m.s. transverse momentum, from the derivative
$\partial F_\pi /\partial Q^2 $ at $Q^2=0$ which can be obtained
through eq. (\ref{eq:drell-yan}). Results obtained that way are consistent
with our ones.}. Lepage et al. \cite{Lep:83}, examining
the process $\pi^0 \to \gamma \gamma$, derived
yet another constraint on the pion wave function:
\beq
\label{eq:pi-gammagamma}
1= \frac{{f_\pi}^2}{12} \int dx_1 \phi(x_1) \Sigma(x_1,\vec k_\perp =0).
\eeq
For the $k_\perp$-dependence of the wave function we assume a simple
Gaussian
\beq
\label{eq:Sigma}
\Sigma(x_1,\vec k_\perp)=16 \pi^2 \beta^2 \,g(x_1)
\,\exp \left( -g(x_1) \beta^2 k_\perp^2 \right),
\eeq
$g(x_1)$ being either  $1$ or $1/x_1 x_2$. In the first case the
wave function $\Psi_0$ factorizes in $x_1$ and $k_\perp$, which
manifestly breaks rotational invariance. This theoretical deficiency
is unessential to the purpose of this article. The
second, non-factorizing case is obtained from a harmonic oscillator rest
frame wave function by equating the energy propagators in both the
rest frame and the infinite momentum frame \cite{Lep:83}. This
particular $k_\perp$-dependence goes along with a factor
$\exp \left[ -\beta^2 m_q^2/x_1 x_2 \right] $ in the DA. Here, $m_q$ is
a constituent quark mass for which we choose a value
of $330 \,{\rm MeV}$. The Gaussian (\ref{eq:Sigma}) is consistent
with the required large-$k_\perp$ behaviour of a soft wave function.
 For the DAs we try two versions, the asymptotic form $\sim x_1 x_2$ and
the CZ form $\sim x_1 x_2 (x_1-x_2)^2$ \cite{Che:82,Hua:91}. Thus
all together
we utilize four examples of pion wave functions:
\newcounter{subequation}
\renewcommand{\theequation}{\arabic{equation}\alph{subequation}}
\begin{eqnarray}
\label{eq:DAs}
\phi_a(x) &=& A_a \, x_1 x_2 \hspace{6.82cm}
g=1
\stepcounter{subequation}\\
\addtocounter{equation}{-1}
\phi_b(x) &=& A_b \, x_1 x_2 (x_1-x_2)^2 \hspace{5.06cm}
g=1
\stepcounter{subequation}\\
\addtocounter{equation}{-1}
\phi_c(x) &=& A_c \, x_1 x_2 \,
\exp\left[ -\frac{{\beta_c}^2 m_q^2}{x_1 x_2}\right]  \hspace{4.05cm}
g=\frac{1}{x_1 x_2}
\stepcounter{subequation}\\
\addtocounter{equation}{-1}
\phi_d(x) &=& A_d \, x_1 x_2 (x_1-x_2)^2 \,
\exp\left[ -\frac{{\beta_d}^2 m_q^2}{x_1 x_2}\right]  \hspace{2.2cm}
g=\frac{1}{x_1 x_2}
\stepcounter{subequation}
\end{eqnarray}
\renewcommand{\theequation}{\arabic{equation}}
Example b) emphasizes most the end-point regions, example c) least of
all. In principle the DAs depend on the renormalization
scale. Asymptotically, that
is for $Q \to \infty$, they all evolve into the asymptotic
form (\ref{eq:DAs}a), which itself
has no evolution. We are going to ignore
the evolution in the main part of this article since it is a
complication of minor importance. We will
however comment below on the effects the evolution may
cause. Finally, a remark is in order concerning the
DAs (\ref{eq:DAs}c) and (\ref{eq:DAs}d). Such functions have been proposed
by Chernyak and Zhitnitsky \cite{Che:82} on the
basis of the two lowest moments of the
pion DA as derived from QCD sum rules.  The CZ ansatz
for the DAs is the subject of
considerable controversy in which we do not want to enter
\footnote{For instance, Braun and Filianov \cite{Bra:89}, also
employing QCD sum rule techniques, found a value of $1.2 \pm 0.3$ for
the pion DA at $x=1/2$, in clear contradiction with the DAs of the
CZ type. It has been pointed out
by Radyushkin \cite{Rad:91} that
the finite size of the vacuum
fluctuations had not been considered in \cite{Che:82}. Taking
these finite-size
effects into account, one finds values for the moments of the
pion DA that are much closer to those
obtained from the asymptotic DA than
the values quoted in \cite{Che:82}.}. We
merely consider (\ref{eq:DAs}c) and (\ref{eq:DAs}d) as examples
of DAs strongly concentrated in the end-point regions.\\
Our wave functions have one free parameter, the oscillator
parameter $\beta$, which we fix by requiring specific values for the
r.m.s. transverse momentum, namely
$350$ and $250\,{\rm MeV}$. In table 1 we
compile the properties of our wave functions, namely the constants $A$,
the oscillator parameters $\beta$, the probabilities of the $q\bar q$
Fock state $P_{q \bar q}$ and the value of the right-hand side of
eq. (\ref{eq:pi-gammagamma}), the $\pi \to \gamma \gamma$
constraint. The value of $350\,{\rm MeV}$ for
the r.m.s. transverse momentum
turns out to be a reasonable value on all accounts. The corresponding
radius is a little smaller than the measured charge radius of the
pion and the constraint (\ref{eq:pi-gammagamma}) is well
satisfied. The probabilities of the $q\bar q$ Fock state also have
plausible values. On the other hand, the value of $250\,{\rm MeV}$ for
the r.m.s. transverse momentum is unrealistic. The radius is too large
and (\ref{eq:pi-gammagamma}) is strongly violated. We, however, keep
this value for comparison.\\ \\
\underline{\it Numerical results:}
The Fourier transform of the $k_\perp$-dependent part of
the wave function reads
\beq
\label{eq:ft-gaussian}
\hat \Sigma (x_1,\vec b)=
4\pi\,\exp\left(-\frac{b^2}{4 g(x_1) \beta^2}\right).
\eeq
Li and Sterman \cite{LiS:92} assume that the dominant $b$-dependence
of the integrand in eq. (\ref{eq:ft-Fpi}) arises from the Sudakov
factor and that the Gaussian in $\hat\Sigma$ can consequently be
replaced by 1. In order to examine that assumption, we
compare in Fig. 1 the Gaussian with the Sudakov factor
[see (\ref{eq:sudakov})]. We only display the case $g=1$
since the other case, $g=1/x_1x_2$, is very similar. Obviously the
intrinsic  $k_\perp$-dependence of the wave function cannot be ignored. For
momentum transfers of the order of a few GeV the wave function damps the
integrand in (\ref{eq:ft-Fpi}) more than the Sudakov factor. Only
at very large values of Q does the Sudakov factor take over. What is
more important at
a given value of $Q$, the Sudakov factor or the wave function depends
on the value chosen for $\Lambda_{QCD}$ and
$\langle {k_\perp}^2 \rangle^{1/2}$. For
the favoured values, $200\,{\rm MeV}$ for
$\Lambda_{QCD}$ and $350\,{\rm MeV}$ for the r.m.s. transverse momentum, the
wave function plays the major r\^ole in the large $b$ behaviour
of the integrand in (\ref{eq:ft-Fpi}) for $Q$ less
than $20\,{\rm GeV}$, whereas
the Sudakov factor begins to dominate beyond that value, which
is well above the experimentally accessible momentum
transfer region.\\
Numerical evaluations of the pion form factor
through (\ref{eq:ft-Fpi}), using the wave functions given
in (\ref{eq:DAs}), confirm the observations made in Fig. 1. For
the two extreme DAs, the one that is most concentrated
in the end-point regions, (\ref{eq:DAs}b), and the one that is
least, (\ref{eq:DAs}c), we display the results in Fig. 2. The
results for the other two examples, (\ref{eq:DAs}a) and
(\ref{eq:DAs}d), lie in between the two extreme cases.
As compared with
the hard-scattering result, eq. (\ref{eq:hs-Fpi}), the inclusion
of the Sudakov factor leads to some suppressions as already
noticed in \cite{LiS:92}. The
intrinsic $k_\perp$-dependence of the
wave functions provides additional suppression
which amounts to about $50\%$, in
the case (\ref{eq:DAs}b), at $Q=2\,{\rm GeV}$  and is
still substantial
at $7\, {\rm GeV}$. As expected the total suppression
obtained from both
the Sudakov factor and the intrinsic $k_\perp$-dependence is
stronger for the wave function (\ref{eq:DAs}b) than for
(\ref{eq:DAs}c). One may regard the Sudakov factor as a part
of the wave function, in fact as its large-$b$ behaviour. In
this case one would not consider the product of the two
functions, $\exp\left[-S\right]$ and $\hat\Sigma$. Rather, to an
admittedly crude approximation, one would take either the Gaussian
in $\hat\Sigma$ or
the Sudakov factor, whichever is the smaller at a given value
of $b$. Fortunately this recipe yields only minor differences
with the first case.\\
The strong suppression due to the intrinsic
$k_\perp$-dependence of the wave function reveals that
the approximation made
in the derivation of the hard-scattering
formula (\ref{eq:hs-Fpi}), namely the
neglect of the $k_\perp$-dependence
in the hard-scattering amplitude (\ref{eq:hs-amplitude}), is invalid
at small momentum transfer. The reason for
this can easily be understood. Using, for instance, a wave function
of the type (\ref{eq:wvfct},\ref{eq:Sigma}) with $g=1$ and, for
the sake of the argument, replacing $\alpha_s$ in (\ref{eq:hs-amplitude}) by
an average value, one can convert (\ref{eq:pre-hs-Fpi}) into
\beq
\label{eq:kt-integrated}
Q^2 F_\pi(Q^2)=\frac{2\pi}{3} {f_\pi}^2 \bar\alpha_s C_F
\int_0^u dz \, e^z \, E_1(z)
\int_z^u \frac{dy}{y} \phi(y/u) \phi(z/y),
\eeq
where $u=Q^2\beta^2/2$ and $E_1$ is the exponential integral. The
dominant contribution comes from the upper limit of the
$z$ integration. Replacing $E_1$ by the leading term of its asymptotic
expansion
\beq
\label{eq:asy-expansion}
z\,e^z\,E_1(z)\sim 1-\frac{1}{z}+\frac{2}{z^2}\mp \ldots ,
\eeq
one obtains the hard-scattering contribution (\ref{eq:hs-Fpi}). But
the higher-order terms provide substantial corrections to it. The
wave function (\ref{eq:DAs}a), for instance, leads to
\beq
Q^2 F_\pi(Q^2) = Q^2 {F_\pi}^{HSP} (Q^2)
\left[ 1-48 \frac{\langle k_\perp^2 \rangle}{Q^2}+ \ldots \right].
\eeq
In Fig. 2 we have also shown the available data for the
pion form factor at large momentum transfer \cite{Beb:76}. This
is not meant as a serious comparison between data and theory
since, as we mentioned above, the data
may suffer from large systematic errors \cite{Car:90}.
Nevertheless, it seems to us that the perturbative contribution
(\ref{eq:ft-Fpi}), including the Sudakov corrections and the intrinsic
$k_\perp$-dependence of the wave function, is too small with
respect to the data, perhaps by a factor of 2 or so. One
may suspect the QCD evolution of the DAs, ignored by us up to now,
to be responsible for that discrepancy. Let us estimate whether this
explanation holds or not. The evolution equation for the pion DA
has been derived in \cite{Lep:80}. For
our example (\ref{eq:DAs}b) it reads
\beq
\label{eq:evolution}
\phi_b(x_1,w)=\phi_a(x_1) \left[1+\frac{2}{3}\, C_2^{3/2}(x_1-x_2)
\left[ \frac{\alpha_s(w^2)}{\alpha_s(\mu_0^2)}\right]^{50/81}\right],
\eeq
where the $C_n^{3/2}(\zeta)$ are the Gegenbauer polynomials, which
are the eigensolutions of the evolution equation. According
to \cite{Li:92} the evolution
parameter $w$ is taken to be $1/b$. Inserting (\ref{eq:evolution})
into eq. (\ref{eq:ft-Fpi}) and
using $\mu_0=0.5$ GeV \cite{Che:82}, we re-evaluate the pion form factor
and find that the evolution reduces even further the predictions
for $F_{\pi}$ (by about $30\%$ for $Q$ between 3 and 4 GeV).
Similar results are obtained for the other examples of wave functions,
(\ref{eq:DAs}c) and (\ref{eq:DAs}d). This
is in agreement with the findings reported in
\cite{Hua:91}.\\
Thus we are tempted to suppose that higher-twist
contributions, besides the $k_\perp$-dependence, are quite strong
in the few GeV region. Sources of such higher twists are contributions
from higher Fock states, $L\neq 0$ components in
the valence Fock state wave
function or genuine soft contributions.\\
An example of genuine soft contributions is provided by the starting point
of our considerations, namely the Drell-Yan
formula (\ref{eq:drell-yan}). Up to now we have only considered those
contributions from (\ref{eq:drell-yan}) which arise
from the perturbative tail of the
wave functions. The size of the ignored soft contributions
until now
can easily be estimated for our wave functions (\ref{eq:DAs}). One
finds
\beq
\label{eq:Fpi-soft}
{F_\pi}^{soft}= \frac{\pi^2}{3}\,{f_\pi}^2\,\beta^2
\int dx_1 \,g(x_1)\,\phi^2(x_1)\,
\exp\left( -g(x_1)\,\beta^2\, {x_2}^2\, Q^2/2 \right).
\eeq
The integral is dominated by the region near $x_1=1$. Other $x_1$ regions
are strongly damped by the Gaussian. Hence $F_\pi$ sensitively reacts
to the behaviour of the DA for $x_1 \to 1$. Numerical evaluations
of (\ref{eq:Fpi-soft}) for our wave functions (\ref{eq:DAs})
show that the examples (\ref{eq:DAs}a)
and (\ref{eq:DAs}c) provide soft contributions
of the right magnitude to fill
in the gap between the perturbative contribution (\ref{eq:ft-Fpi}) and the
experimental data
\footnote{The Drell-Yan formula (\ref{eq:drell-yan}) has
been obtained by a transverse boost
to the infinite-momentum frame. Since we know the hadronic wave
function only approximately, it may happen that other boost
directions provide different results for the pion form factor. For
a discussion of such eventual violations of rotational
invariance, we refer to \cite{Isg:89,Saw:92}.
} ($Q^2\,F_\pi^{soft} \simeq 0.25\,{\rm GeV^2}$
at $Q=2\,{\rm GeV}$). As required by
the consistency of the entire picture, $F_\pi^{soft}$
decreases faster with increasing $Q$ than the perturbative contribution.
The exponential $\exp(-\beta^2 m_q^2/x_1 x_2)$ in the example (\ref{eq:DAs}c)
turns out to be quite important since it is effective in the end-point
regions. It reduces the size of $Q^2\,F_\pi^{soft}$ substantially and alters
its asymptotic behaviour from a $1/Q^2$ behaviour to an exponentially
damped decrease; $Q^2\,F_\pi^{soft}$ becomes equal
to the perturbative contribution at $Q\simeq 5\,{\rm GeV}$. The
soft contributions
are also subject to Sudakov corrections. An estimate of these corrections
on the basis of eq. (\ref{eq:sudakov}) reveals, however, that
they only amount to
a few per cent for the examples (\ref{eq:DAs}a)
and (\ref{eq:DAs}c). The wave functions
(\ref{eq:DAs}b) and (\ref{eq:DAs}d) provide very
large soft contributions because of
their strong concentration in the end-point regions. Therefore, these
examples appear to be unrealistic
[see also the discussion after eq. (\ref{eq:DAs})]. Soft
contributions of the type (\ref{eq:Fpi-soft}) have
extensively been discussed
in \cite{Isg:89}. Our results for $F_\pi^{soft}$ are similar in trend, but
smaller in size than those presented in \cite{Isg:89}. There
are three reasons
for the differences in the size of the soft contributions:
Isgur et al. \cite{Isg:89} assume that $P_{q\bar q}=1$, they neglect the
exponential $\exp(-\beta^2 m_q^2/x_1x_2)$ and utilize DAs that are
proportional to $\sqrt{x_1x_2}$ instead of $\sim x_1x_2$. Strong
soft contributions to the pion form factor have also
been obtained with QCD sum rules \cite{Rad:91}.
Finally we mention $L\neq 0$ components of the wave function as
another source of higher-twist contributions. For instance, the
valence Fock state component of the pion may be expressed by
\beq
\int\frac{dx_1\,d^{\;\!2}k_\perp}{16\pi^3}
\frac{1}{\sqrt{2}}\, \left( \sla{p}  +m \right)\,
\left[\Psi^0(x_1,\vec k_\perp)
+ \sla{k}_\perp \Psi^1(x_1,\vec k_\perp)\right]\, \gamma_5 ,
\eeq
where $p$ denotes the pion's momentum and $m$ its mass. Ralston
and Pire \cite{Ral:92} consider an even more general ansatz
for the valence Fock state wave function of the pion. Thus one  may have
additional $k_\perp$-dependent contributions to the pion form
factor, which are not necessarily small. Perhaps one
obtains a better description of the pion form factor
with such contributions although at the expense of the introduction
of a new, a priori unknown, phenomenological function. Contributions
from higher Fock states suffer from the same disadvantage; also
in this case new phenomenological functions have to be introduced.
Such extensions of the hard-scattering model are certainly not
very attractive but are, perhaps, necessary.\\ \\
\underline{\it Summary:} On the basis of our numerical studies we
conclude that the intrinsic $k_\perp$-dependence of the wave
function has to be taken into account for a reliable quantitative
estimate of the perturbative QCD contribution to the pion
form factor. We are aware that this introduces a new phenomenological
element into the calculation. The disadvantage is, at least
partially, compensated by the fact that the inclusion of the
intrinsic $k_\perp$-dependence renders the perturbative contribution
even more self-consistent than the Sudakov suppression already
does. Applying the criterion of self-consistency as suggested
by Li and Sterman \cite{LiS:92} (see above), we can conclude that
perturbative QCD begins to be self-consistent for $Q$ between
1 and 2 GeV (for $\langle k_\perp^2 \rangle^{1/2}=350\,{\rm MeV}$). The value
for $Q$ at which self-consistency sets in depends on the wave
function. It is larger for the end-point concentrated wave
functions (\ref{eq:DAs}b,d) than for the other examples
(\ref{eq:DAs}a,c). However, the perturbative
contribution, although self-consistent, is presumably too small with
respect to the data. It thus seems that other contributions (higher
twists) also play an important r\^ole in the few GeV region. We
see no argument why our observations should not apply as well
to other exclusive observables, such as the nucleon form factor.\\
\newpage
\begin{table}[t]
\label{table:parameters}
\caption{Properties of the pion wave functions (20).}
\vspace{.5 cm}
\centering
\begin{tabular}{|c||c|c|c|c||c|c|c|c|}\hline
\rule[-3mm]{0mm}{8mm}
DA & $A$ & $\beta^2 [{\rm GeV}^{-2}]$ & $P_{\bar q q}$
   & R.h.s. of (\ref{eq:pi-gammagamma})
   & $A$ & $\beta^2 [{\rm GeV}^{-2}]$ & $P_{\bar q q}$
   & R.h.s. of (\ref{eq:pi-gammagamma})\\
\hline\hline
\rule[-3mm]{0mm}{8mm}
a) & 6      & 4.082 & 0.284 & 0.947
   & 6      & 8.0   & 0.557 & 1.856 \\
\rule[-3mm]{0mm}{8mm}
b) & 30     & 4.082 & 0.338 & 0.947
   & 30     & 8.0   & 0.663 & 1.856 \\
\rule[-3mm]{0mm}{8mm}
c) & 10.07  & 0.883 & 0.320 & 1.044
   & 16.41  & 1.787 & 0.681 & 2.006 \\
\rule[-3mm]{0mm}{8mm}
d) & 50.05  & 0.571 & 0.357 & 0.983
   & 83.59  & 1.247 & 0.721 & 1.915 \\
\hline\hline
\rule[-3mm]{0mm}{8mm}
&\multicolumn{4}{c||}{$\sqrt{\langle k_T^2 \rangle}=350\,{\rm MeV}$}
&\multicolumn{4}{c|}{$\sqrt{\langle k_T^2 \rangle}=250\,{\rm MeV}$}\\
\hline
\end{tabular}
\vspace{3 cm}
\end{table}
{\Large\bf Figure captions}
\begin{description}
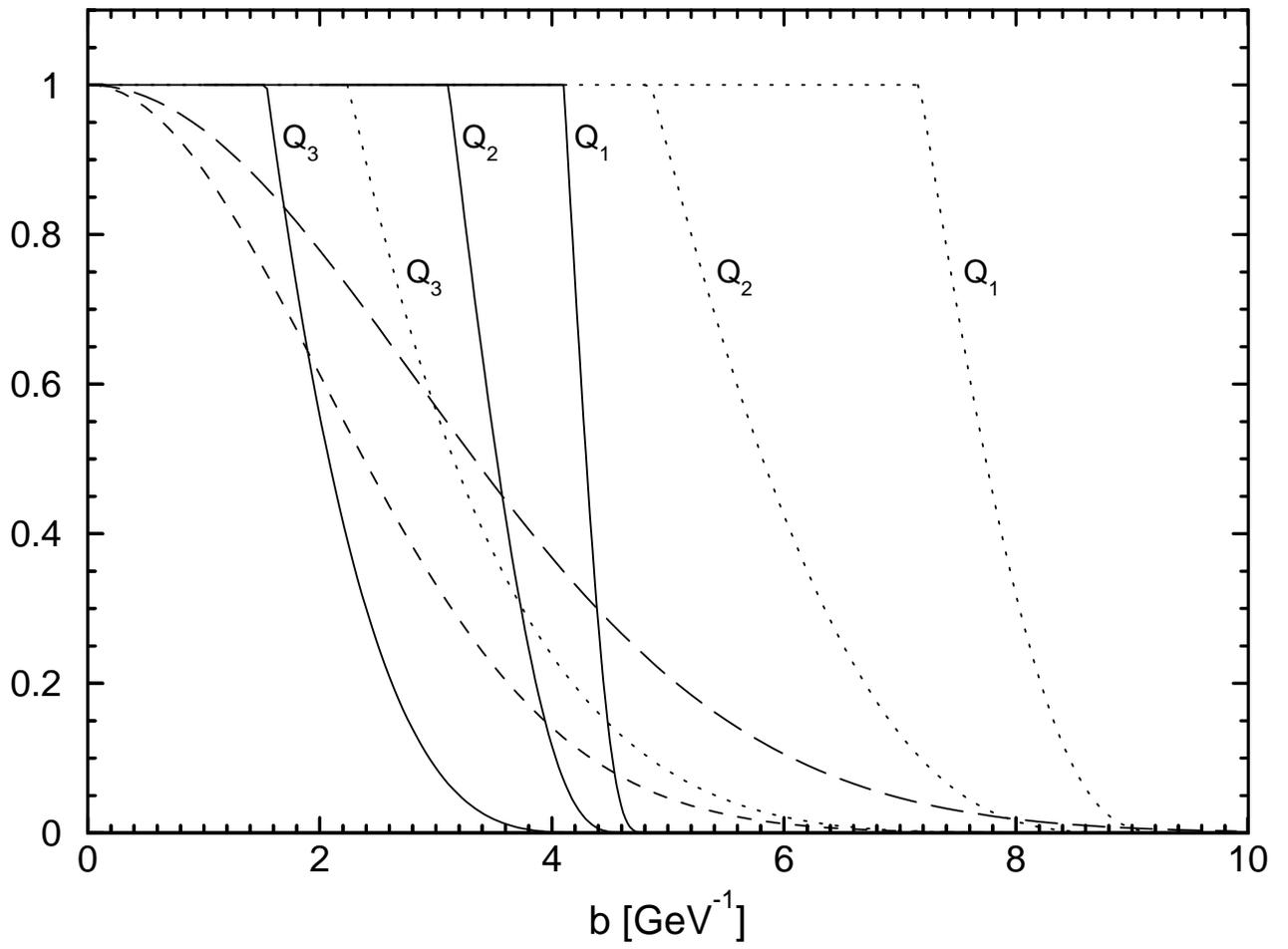
\item[Fig. 1:]The Sudakov factor, evaluated at $x_1=y_1=1/2$, and
the Gaussian $\exp\left(-b^2/2 \beta^2\right)$ [see
eq. (\ref{eq:ft-gaussian})] as functions of the transverse
separation $b$. The Gaussian is shown for a r.m.s. transverse
momentum of $350\,{\rm MeV}$ (short-dashed line)
and $250\,{\rm MeV}$ (long-dashed line). The Sudakov
factor is evaluated for $\Lambda_{QCD}=200\,{\rm MeV}$
(solid lines) and $100\,{\rm MeV}$ (dotted lines)
for $Q_1=2\,{\rm GeV}$, $Q_2=5\,{\rm GeV}$
and $Q_3=20\,{\rm GeV}$ ($x_1=x_2=1/2$).
\item[Fig. 2a:] The pion form factor as a function of $Q^2$ evaluated
with the wave function (\ref{eq:DAs}b) and
$\Lambda_{QCD}=200\,{\rm MeV}$.
The dash-dotted line is obtained from the hard-scattering formula
with an $\alpha_s$ cut-off at 0.5 and the dashed line
from (\ref{eq:ft-Fpi}) ignoring the intrinsic
$k_\perp$-dependence. The solid and the dotted
lines represent the complete result obtained from (\ref{eq:ft-Fpi})
taking into account both the Sudakov factor and the intrinsic
$k_\perp$-dependence for $\langle k_\perp^2 \rangle =350\,{\rm MeV}$ and
$250\,{\rm MeV}$, respectively. Data
are taken from \cite{Beb:76} ($\circ$ 1976, $\bullet$ 1978).
\item[Fig. 2b:]As Fig. 2a but using the
wave function (\ref{eq:DAs}c). Note the modified scale of
the abscissa.
\end{description}
\newpage
\bfig[h]
\epsfbox{fig1.psc}
{\bf \Large Fig. 1}
\efig
\bfig[h]
\epsfbox{fig2a.psc}
{\bf \Large Fig. 2a}
\efig
\bfig[h]
\epsfbox{fig2b.psc}
{\bf \Large Fig. 2b}
\efig
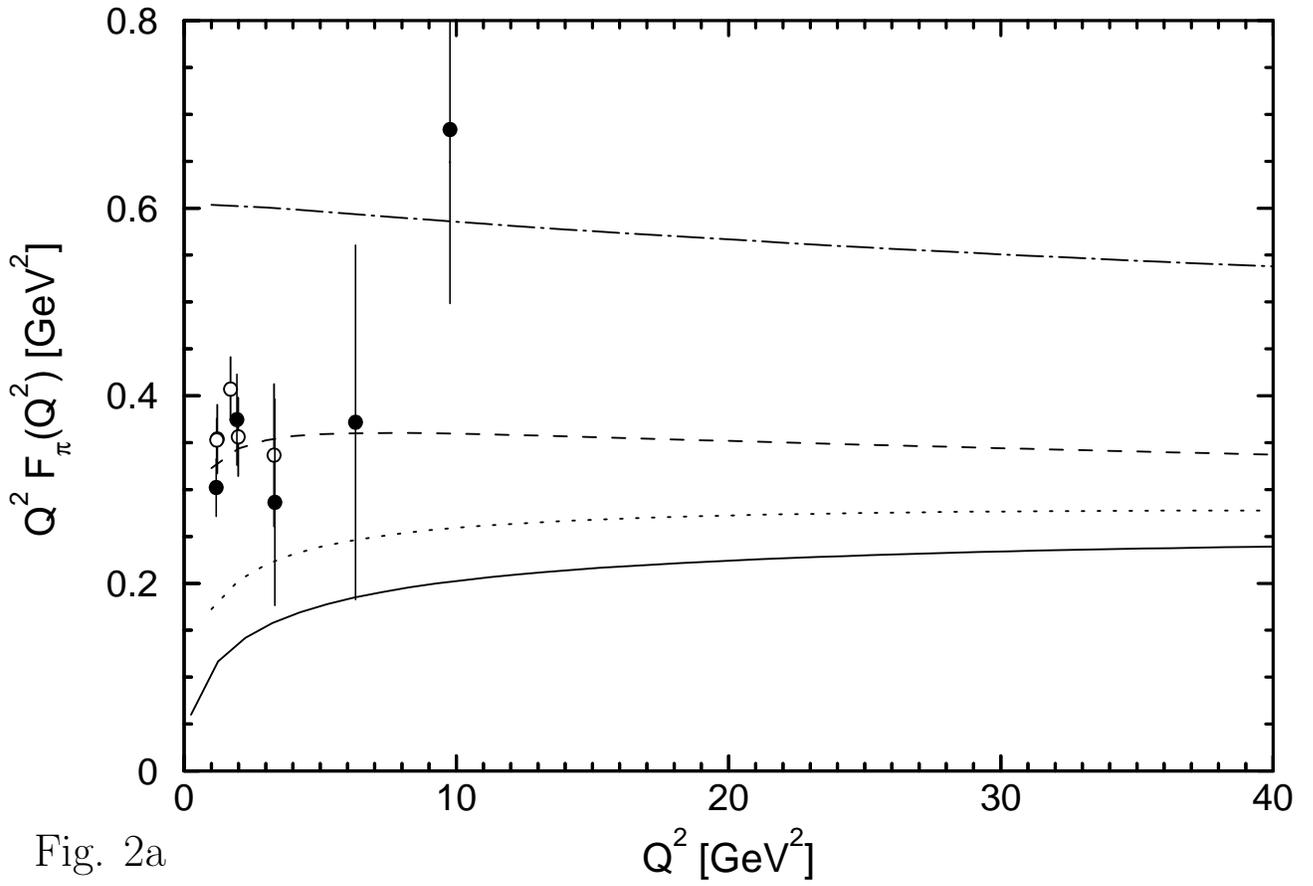
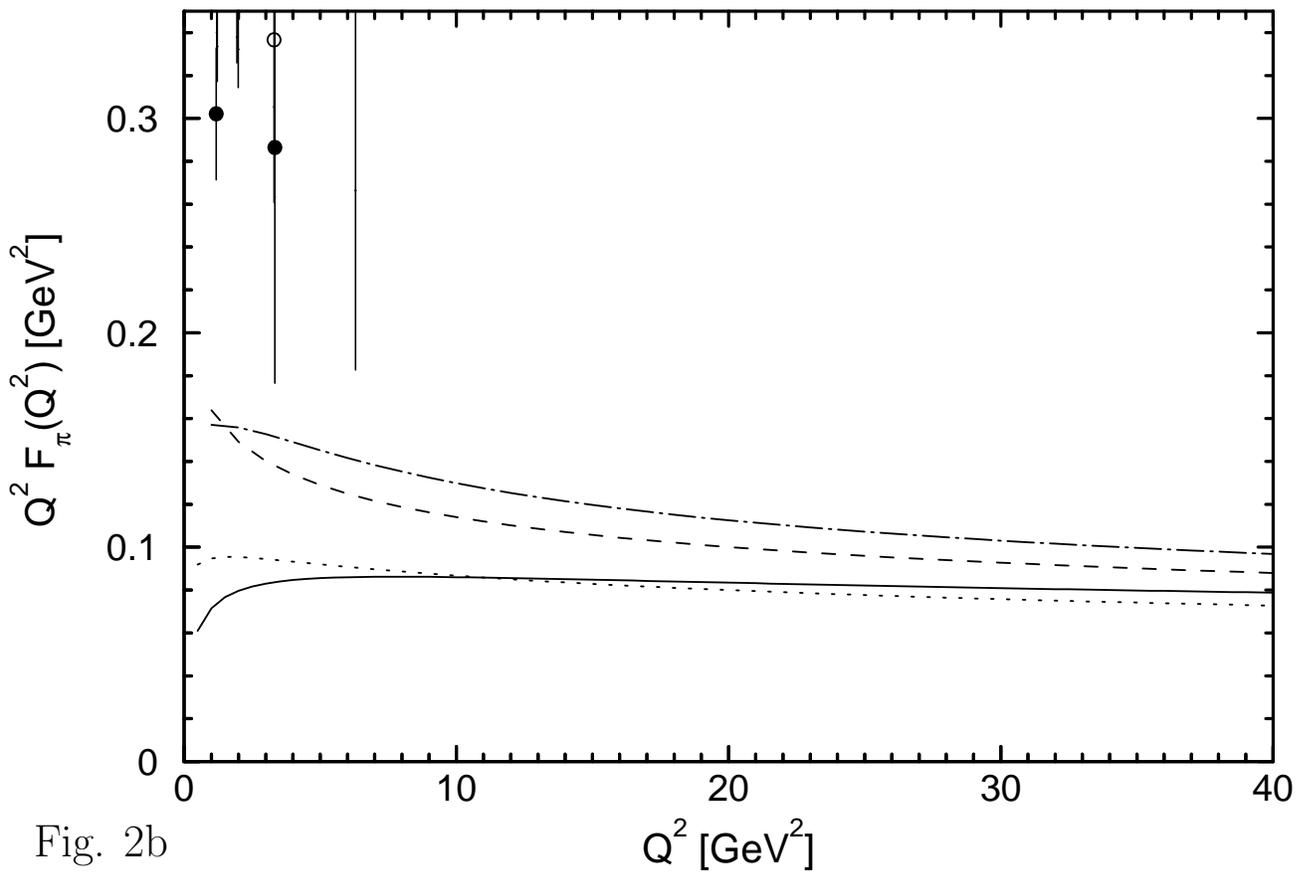
\end{document}